\documentclass[conference]{IEEEtran}
\IEEEoverridecommandlockouts
\usepackage{cite}
\usepackage{amsmath,amssymb,amsfonts}
\usepackage{algorithmic}
\usepackage{graphicx}
\usepackage{textcomp}
\usepackage{xcolor}
\usepackage{url}

\def\BibTeX{{\rm B\kern-.05em{\sc i\kern-.025em b}\kern-.08em
    T\kern-.1667em\lower.7ex\hbox{E}\kern-.125emX}}

\newtheorem{thm}{Theorem}

\newtheorem{definition}{Definition}
\newtheorem{lemma}{Lemma}

\newtheorem{remark}{Remark}
\usepackage{bbm}
\newcommand{\M}{\mathsf{M}}
\newcommand{\m}{\mathsf{m}}

\newcommand{\mw}[1]{{\color{black}#1}}

\begin{document}
\interdisplaylinepenalty=0
\title{Cooperative Multi-Sensor Detection under Variable-Length Coding\\
}
\author{\IEEEauthorblockN{Mustapha Hamad}
	\IEEEauthorblockA{\textit{LTCI, Telecom Paris, IP Paris} \\
		91120 Palaiseau, France\\
		mustapha.hamad@telecom-paris.fr}
	\and
	\IEEEauthorblockN{Michèle Wigger}
	\IEEEauthorblockA{\textit{LTCI, Telecom Paris, IP Paris} \\
		91120 Palaiseau, France\\
		michele.wigger@telecom-paris.fr}
			\and
	\IEEEauthorblockN{Mireille Sarkiss}
	\IEEEauthorblockA{\textit{SAMOVAR, Telecom SudParis, IP Paris} \\
		91011 Evry, France\\
		mireille.sarkiss@telecom-sudparis.eu}
}
\allowdisplaybreaks[4]
\sloppy
\maketitle

\begin{abstract}
We investigate the testing-against-independence problem \mw{over a cooperative MAC}  with two sensors and a single detector under an average rate constraint on the sensors-detector links. For this setup, we design a variable-length coding scheme that maximizes the achievable type-II error exponent when the type-I error probability is limited to $\epsilon$. Similarly to the single-link result, we show here that the optimal error exponent depends on $\epsilon$ and  that variable-length coding allows to \mw{increase the  rates over the optimal} fixed-length coding \mw{scheme} by the factor $(1-\epsilon)^{-1}$. 

%This  paper  characterizes  the  optimal exponent  for a distributed   hypothesis   testing-against-independence  problem  when  the expected rate  of  the  sensor-detector link is constrained. Unlike for the well-known result  that  holds  under  a maximum rate  constraint  and where a strong converse holds, here the optimal exponent dependson  the  allowed  type-I  error  exponent.  Specifically,  if  the  type I error  probability  is  limited then  the  optimal  type-II  error exponent under an expected rate constraint R coincides with the optimal type-II error exponent under a maximum rate constraint.
\end{abstract}


\section{Introduction}

Motivated by the  broadly emerging Internet of Things (IoT) applications, distributed hypothesis testing problems gained  increasing  attention recently. In such problems, sensors  send information about their observations to one or multiple decision centers. Then, the decision centers attempt to detect the joint distributions underlying the data observed at all the terminals including their own observations.

Our focus is on binary hypothesis testing with a null hypothesis and an alternative hypothesis. We are interested in maximizing the exponential decay (in the number of observed samples) of the probability of error under the alternative hypothesis, given a constraint  on the probability of error under the null hypothesis. The study of such a Stein setup has a long history in the information theoretic literature, see e.g.,  \cite{Ahlswede,Han,Amari,Wagner,Kim,Michele2,SW18,Michele} which study point-to-point, interactive, cascaded, and multi-sensor and/or multi-detector systems.
 %null hypothesis ($\mathcal{H} = 0$) or the alternate hypothesis ($\mathcal{H} = 1$). There are two types of error in such problem: Type-I error probability (probability of guessing $\hat{\mathcal{H}} = 1$ when $\mathcal{H} =0$ is true), labeled as “false positives”, and Type-II error probability (probability of guessing $\hat{\mathcal{H}} = 0$ when $\mathcal{H} =1$ is true), labeled as “missed detection”.
%  A main goal is to decay Type-II error probability exponentially fast to 0, i.e. maximizing the error exponent $\theta$,  while maintaining Type-I error probability within a predefined threshold $\epsilon$. In practical scenarios, this can represent some security measures to detect all possible attacks under a predefined threshold of false alarms
All these works constrain the \emph{maximum} rate of communication  between terminals, and a fixed-length communication scheme is obviously optimal.  Recently,  the authors of \cite{MicheleVL} proposed to only constrain the \emph{average} rate of communication, and they presented a variable-length coding scheme that under this weaker constraint improves the maximum achievable error exponent.   %An average constraint on the communication rate can model for example  practical in IoT applications where sensors are allocated a variable bandwidth under an average link capacity. \cite{MicheleVL} showed potential gains in the optimal error exponent as function of available rate for the point-to-point setup. 
The present work is the first extension of the point-to-point average-rate scenario in \cite{MicheleVL}  and the corresponding variable-length coding scheme to systems with \mw{multiple sensors}. 

\mw{Specifically,} we consider the two-sensors  single-detector system in Fig.~\ref{fig:MAC Cooperative Setup}, where the first sensor communicates over a shared link to the second sensor and the detector, and after receiving this message, also the second sensor communicates with the detector. The two sensors observe the sequences $X_1^n$ and $X_2^n$, respectively, and the detector observes $Y^n$, where we assume \mw{that the following Markov chain holds both under the null hypothesis $\mathcal{H}=0$ as well as under the alternative hypothesis $\mathcal{H}=1$:}
\begin{equation}\label{eq:Mc}
X_1^n \leftrightarrow X_2^n \leftrightarrow Y^n
\end{equation}   We consider the \emph{testing-against-independence} scenario where under the alternative hypothesis $\mathcal{H}=1$ the observations at the two sensors are independent of the observations at the detector. \mw{We further assume that the sensors' observations $X_1^n, X_2^n$ follow the same \emph{joint} distribution and  the decision center's observation $Y^n$ follows the same \emph{marginal} distribution under both hypotheses}. A more general version of our problem \mw{(without  Markov chain  \eqref{eq:Mc})} was studied in \cite{zhao2018distributed}, but under a maximum rate constraint. 

In this paper, we characterize the maximum achievable error exponent $\theta_{\epsilon}^*(R_1, R_2)$ under the alternative hypothesis when the error probability under the null hypothesis is not allowed to exceed $\epsilon$, and where here $R_1$ and $R_2$ denote the rates of communication from the first and the second sensors, respectively. As we show in this paper, and in contrast to the  optimal  error exponent under a maximum rate constraint $\theta^*_{\epsilon,\textnormal{Fix}}(R_1,R_2)$  \cite{zhao2018distributed},\footnote{In the converse proof of \cite[Theorem 2]{zhao2018distributed}, the second step used to upper bound the rate $R_1$ relies on  the Markov chain $X_{1i}\leftrightarrow(M_1,X_1^{i-1}X_2^{i-1})\leftrightarrow X_{2_{i+1}}^{n}$, which does not necessarily hold. The result of \cite{zhao2018distributed} remains however valid under the Markov chain \eqref{eq:Mc}, see Remark~\ref{rem:fixed} ahead.} the optimal exponent $\theta_{\epsilon}^*(R_1, R_2)$ depends on $\epsilon$. In fact, as a main result, we obtain
\begin{equation}\theta_{\epsilon}^*(R_1,R_2)= \theta^*_{\epsilon,\textnormal{Fix}}(R_1/(1-\epsilon), R_2/(1-\epsilon)). 
\end{equation}
Thus, through variable-length coding we can increase all available rates in the network  by the factor $(1-\epsilon)^{-1}$. A similar observation was already made for the point-to-point setup studied in \cite{MicheleVL}. In this sense, the current paper extends the conclusion to multiple links, \mw{and it shows in particular that the rate-increase can be attained on all links simultaneously.}
	%	Similarly, the required Markov chain $X_{1i}\leftrightarrow(M_1,X_1^{i-1}X_2^{i-1})\leftrightarrow X_{2_{i+1}}^{n}$ second equality  More specifically, they assume that $I(M_1;X_1^nX_2^n) = \sum_{i=1}^{n}{I(M_1X_1^{i-1}X_{2_{(i+1)}}^{n};X_{1i}X_{2i})}$ and $I(M_2;X_1^nX_2^nY^n|M_1) = \sum_{i=1}^{n}{I(M_2;X_{1i}X_{2i}Y_i|M_1X_1^{i-1}X_{2_{(i+1)}}^{n}Y^{i-1})}$ which is not true and can be refuted with the following example: Assume

\mw{\textit{Notation:}}
We follow the notation in \cite{ElGamal} and \cite{MicheleVL}. In particular,  \mw{we use sans serif font for bit-strings: e.g., $\m$ for a deterministic and $\M$ for a random bit-string. We let  $\mathrm{string}(m)$ denote the shortest bit-string representation of a positive integer  $m$, and for any bit-string $\m$ we let  $\mathrm{len}(\m)$  and $\mathrm{dec}(\m)$ denote its length and its corresponding positive integer}. %, and the function $\mathrm{str}(m)$ maps a decimal non-number into its shortest bit-string representation. 
We use  $h_{\textnormal{b}}(\cdot)$ for the binary entropy function.

\section{System Model}

\begin{figure}[htbp]
	\centerline{\includegraphics[width=8.5cm, scale=0.86]{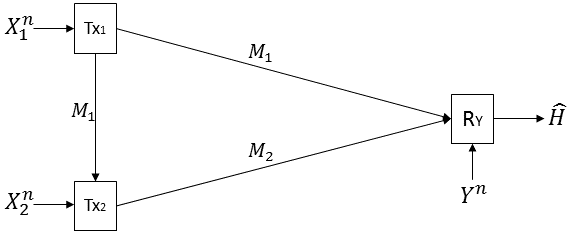}}
	\caption{Cooperative MAC Setup with 2 transmitters and 1 receiver.}
	\label{fig:MAC Cooperative Setup}
\end{figure}

Consider the distributed hypothesis testing problem in Fig.~\ref{fig:MAC Cooperative Setup} in the special case of testing against independence where
\mw{\begin{IEEEeqnarray}{rCl}
& &\textnormal{under } \mathcal{H} = 0: (X_1^n,X_2^n,Y^n) \sim \textnormal{i.i.d.} \, P_{X_1X_2}\cdot P_{Y|X_2} ;\IEEEeqnarraynumspace\\
%& &\textnormal{and  } \nonumber\\
& &\textnormal{under } \mathcal{H} = 1: (X_1^n,X_2^n,Y^n) \sim \textnormal{i.i.d.} \, P_{X_1X_2}\cdot P_{Y}.
\end{IEEEeqnarray}}
\vspace{-0.85cm}
\vspace{-0.1cm}

Specifically, the system consists of two transmitters ($T_{X_1}$ and $T_{X_2}$) and a receiver ($R_Y$). $T_{X_1}$ observes the source sequence $X_1^n$ and sends its bit-string message $\M_1 = \phi_1^{(n)}(X_1^n)$ to both $T_{X_2}$ and $R_Y$, where the encoding function is of the form $\phi_1^{(n)} : \mathcal{X}_1^n \to \{0,1\}^{\star}$ \mw{and satisfies} the rate constraint
\begin{equation}\label{eq:Rate1}
\mathbb{E}\left[\mathrm{len}\left(\M_1\right)\right]\leq nR_1.
\end{equation} $T_{X_2}$ observes the source sequence $X_2^n$ and with the message $\M_1$ received from $T_{X_1}$, it computes \mw{the bit-string} message $\M_2 = \phi_2^{(n)}\left(X_2^n,\M_1\right)$ using some encoding function $\phi_2^{(n)} : \mathcal{X}_2^n \times \{0,1\}^{\star} \to \{0,1\}^{\star}$ satisfying the rate constraint
\begin{equation}\label{eq:Rate2}
\mathbb{E}\left[\mathrm{len}\left(\M_2\right)\right]\leq nR_2.
\end{equation} $T_{X_2}$ sends message $\M_2$ to $R_Y$ which decides on the hypothesis $\mathcal{H}=\{0,1\}$ based on the messages $\M_1$ and $\M_2$ and its own observation $Y^n$. That means, using a decoding function $g^{(n)} : \mathcal{Y}^n \times \{0,1\}^{\star} \times \{0,1\}^{\star} \to \{0,1\}$, it produces:
\begin{equation}
\hat{\mathcal{H}} = g^{(n)}\left(\M_1,\M_2,Y^n\right) \;   \in\{0,1\}.
\end{equation}
%
%In order for the decoder, which receives two messages $m_1$ and $m_2$, to take a decision in favor of one of the two hypotheses, we define the acceptance region as: \begin{equation}
%{\mathcal{A}}_n \triangleq \{\left(m_1,m_2,y^n\right): g^{n}\left(m_1,m_2,y^n)\right)=0\}
%\end{equation}

The  goal is to design encoding and decision functions such that their type-I error probability  
\begin{equation}
\alpha_n \triangleq \Pr[\hat{\mathcal{H}} = 1|\mathcal{H}=0]
\end{equation}
\mw{stays below} a given threshold and the  type-II error probability
\begin{equation}
\beta_n \triangleq \Pr[\hat{\mathcal{H}} = 0|\mathcal{H}=1]
\end{equation}
decays to 0 exponentially fast.

\begin{definition} \mw{Error exponent $\theta\geq 0$ is called \emph{$\epsilon$-achievable}} if there exists a sequence of encoding and decision functions $\{\phi_1^{(n)},\phi_2^{(n)},g^{(n)}\}$ satisfying
\mw{\begin{eqnarray}\label{type1constraint}
\alpha_n & \leq& \epsilon,\\ 
\label{thetaconstraint}
\varlimsup_{n \to \infty} \inf {1 \over n} \log{1 \over \beta_n} &\geq& \theta.
\end{eqnarray}}
\mw{The supremum over all $\epsilon$-achievable error exponents is called the \emph{optimal error exponent} and is denoted $\theta^*_{\epsilon}(R_1,R_2)$.}
\end{definition}
\begin{remark}
	The present setup differs from the one considered by Zhao and Lai \cite{zhao2018distributed} \mw{only in that \cite{zhao2018distributed} imposes the more stringent constraints 
	\begin{IEEEeqnarray}{rCl}\label{eq:fixed_Rate}
	\mathrm{len}\left(\M_i\right)\leq nR_i, \quad i\in\{1,2\},
	\end{IEEEeqnarray}
instead of
	the \emph{expected rate constraints}  \eqref{eq:Rate1} and \eqref{eq:Rate2}.} Under the rate-constraints \eqref{eq:fixed_Rate}, without loss of optimality, the two transmitters can send messages $\M_1$ and $\M_2$ of fixed lengths. 
\end{remark}
\section{Main Results}
\begin{thm}\label{thm1}
	There exist auxiliary random variables $U_1$ and $U_2$ such that the optimal error exponent is given by:
\begin{equation}
\theta_{\epsilon}^{*}\left(R_1,R_2\right) = \max\limits_{\substack{P_{U_1|X_1}, P_{U_2|U_1X_2}\colon \\R_1 \geq \left(1-\epsilon\right)I\left(U_1;X_1\right)\\R_2 \geq \left(1-\epsilon\right) I\left(U_2;X_2|U_1\right)\\U_1\leftrightarrow X_1\leftrightarrow (X_2,Y)\\U_2\leftrightarrow (X_2,U_1)\leftrightarrow (X_1,Y)}} I\left(U_1U_2;Y\right)
\end{equation}
where mutual information quantities are calculated according to the joint pmf $P_{U_1U_2X_1X_2Y}\!\triangleq\!P_{U_1|X_1}P_{U_2|U_1X_2}P_{X_1X_2}P_{Y|X_2}$.
\end{thm}
\begin{IEEEproof} Achievability is proved in Section~\ref{acheivability} and the converse in Section~\ref{converse}.  
	\end{IEEEproof}
\begin{lemma}\label{lem1}
In Theorem~\ref{thm1}, it suffices to choose $U_1$ and $U_2$ over alphabets of sizes $\vert \mathcal{U}_1 \vert \leq \vert \mathcal{X}_1 \vert +2$ and $\vert \mathcal{U}_2 \vert \leq \vert \mathcal{U}_1 \vert \vert \mathcal{X}_2 \vert + 1$.
\end{lemma}
\begin{IEEEproof}
Omitted. It follows by standard applications of Carath\'eodory's theorem, see \cite[Appendix C]{ElGamal}.
\end{IEEEproof}

\subsection{Comparing Variable-Length with Fixed-Length Coding}
For comparison, we also present the optimal error exponent under fixed-length coding. 
\begin{remark}\label{rem:fixed}
	Under  fixed-length coding, i.e., under rate constraints \eqref{eq:fixed_Rate}, the optimal error exponent $\theta^{*}_{\epsilon,\textnormal{Fix}}\left(R_1,R_2\right) $ is \cite{zhao2018distributed}:
	\begin{equation}
	\theta^{*}_{\epsilon,\textnormal{Fix}}\left(R_1,R_2\right) = \max\limits_{\substack{P_{U_1|X_1}, P_{U_2|U_1X_2}\colon\\R_1 \geq I\left(U_1;X_1\right)\\R_2 \geq I\left(U_2;X_2|U_1\right)\\U_1\leftrightarrow X_1\leftrightarrow (X_2,Y)\\U_2\leftrightarrow (X_2,U_1)\leftrightarrow (X_1,Y)}} I\left(U_1U_2;Y\right)
	\end{equation}
where mutual information quantities are calculated according to the joint pmf $P_{U_1U_2X_1X_2Y}\!\triangleq\!P_{U_1|X_1}P_{U_2|U_1X_2}P_{X_1X_2}P_{Y|X_2}$.
\end{remark}
\begin{IEEEproof}
	Achievability can be proved  as described in Section~\ref{acheivability} when the set $\mathcal{S}_n$ is replaced by an empty set.
	The converse can be shown as in Section~\ref{converse} if inequality \eqref{Miub}, i.e., $H(\tilde{\M}_i) \leq \frac{nR_i}{\Delta_n}   \left(1 + {h_b\left({\Delta_n \over nR_i}\right)}\right)$, is replaced by the trivial inequality $H(\tilde{\M}_i)\leq nR_i$. A more direct proof is also possible, similar to the one in \cite{zhao2018distributed}; the converse proof in \cite{zhao2018distributed} relies however on a wrong Markov chain, \mw{ see  our footnote 1.}
\end{IEEEproof}

\begin{figure}[htbp]
	\centerline{\includegraphics[width=7cm, scale=0.6]{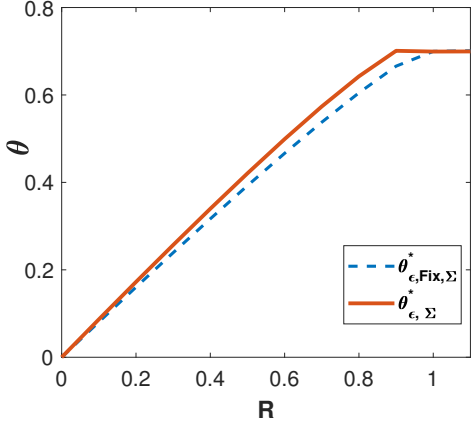}}%\vspace{-0.2cm}%
	\caption{\mw{Optimal exponents under variable-length and fixed-length coding under a sum-rate constraint for above example  with $a = 0.5, p=0.75, q=0.95, \textnormal{and } \epsilon=0.07$.}}
	\label{fig:DSBS Cooperative MAC VL vs FL} 
\end{figure}

We examine the gain provided by variable-length coding \mw{on the cooperative MAC at hand of an example. Let $X_1,S,T$ be  independent Bernoulli random variables of parameters $a,p,q\in[0,1]$ and set $X_2=X_1 \oplus T$ and $Y=X_2 \oplus S$. % for $(S,T)$Consider a doubly-symmetric binary source where  independent %Fig.~\ref{fig:DSBS MAC channel} where $\Pr(X_2=0|X_1=0)=\Pr(X_2=1|X_1=1)=p$ and $\Pr(Y=0|X_2=0)=\Pr(Y=1|X_2=1)=q$ for $p,q \in [0,1]$.} %\vspace{-0.1cm}%
%\begin{figure}[htbp]
%	\centerline{\includegraphics[width=6cm]{figures2/Double_DSBS_Channel}}
%	\caption{Double DSBS Channels}
%	\label{fig:DSBS MAC channel}
%\end{figure}
For this example, Fig.~\ref{fig:DSBS Cooperative MAC VL vs FL} plots  the \mw{optimal error exponents of variable-length and fixed-length coding} \emph{under a sum-rate constraint}:  \begin{eqnarray}
\displaystyle \theta_{\epsilon,\Sigma}^*(R)&:=&\max_{R_1,R_2\geq 0\colon R_1+R_2 \leq R} \theta_{\epsilon}^*(R_1,R_2)\\
\displaystyle \theta_{\epsilon,\textnormal{Fix},\Sigma}^*(R)&:=&\max_{R_1,R_2\geq 0\colon R_1+R_2 \leq R} \theta_{\epsilon,\textnormal{Fix}}^*(R_1,R_2)
\end{eqnarray}  for $\epsilon =0.07$ and in function of the sum-rate $R$.}

Note that the optimal type-II error exponent under an expected rate constraint  $R$ coincides with the optimal type-II error exponent under a maximum rate constraint $(1- \epsilon)R$. \color{black} Moreover, as $R$ increases, both error exponents $\theta_{\epsilon,\Sigma}^*$ and $\theta_{\epsilon,\textnormal{Fix},\Sigma}$ tend to the optimal exponent $I(X_1X_2;Y)$ that can be obtained in a central hypothesis testing problem where the detector directly observes all theses sequences $X_1^n,X_2^n,Y^n$. In particular, both simulated optimal error exponents reach a  value of 0.7011 at $R=1.1$ which is almost 98.25\% of $I(X_1X_2;Y) = 0.7136$.
\color{black}
\vspace{-4mm}

\vspace{-3mm}

\section{Achievability Proof}\label{acheivability}
Fix a large blocklength $n$, a small number $\mu \in (0,\epsilon)$, and conditional pmfs $P_{U_1|X_1}$ and \mw{$P_{U_2|U_1X_2}$} such that:
\begin{equation}
R_1 = \left( 1- \epsilon + \mu \right)\left(I(U_1;X_1) + 2\mu\right)
\end{equation}
\begin{equation}
R_2 = \left( 1- \epsilon + \mu \right)\left(I(U_2;X_2|U_1) + 2\mu\right)
\end{equation}
where mutual information quantities are calculated according to the joint pmf
\begin{equation}
P_{U_1U_2X_1X_2Y} \triangleq P_{U_1|X_1} \cdot P_{U_2|U_1X_2} \cdot P_{X_1X_2} \cdot P_{Y|X_2}.
\end{equation}
\mw{Randomly generate a codebook 
\begin{equation}
C_{U_1} \triangleq \left\{u_1^{n}(m_1): m_1 \in \left\{1,\cdots,2^{n\left(I\left(U_1;X_1\right) + \mu \right)}\right\}\!\right\}
\end{equation}
by drawing all entries i.i.d. according to the marginal pmf $P_{U_1}$.}
For each codeword $u_1^{n}(m_1)$, generate a codebook
\begin{equation}
C_{U_2}(m_1) \triangleq \left\{u_2^{n}(m_2|m_1): m_2\!\in\! \left\{\!1,\!\cdots\!,2^{n\left(I\left(U_2;X_2|U_1\right) + \mu \right)}\right\}\!\right\}\!,
\end{equation}
by drawing the \textit{j}-th entry of each codeword according to the marginal pmf $P_{U_2|U_1}$.
Also, choose a subset $\mathcal{S}_n$ of the typical set $\mathcal{T}_{\mu}^{(n)}(P_{X_1})$ with probability slightly less than $\epsilon$:
\begin{equation}
\mathcal{S}_n \subseteq \mathcal{T}_{\mu}^{(n)}(P_{X_1}) : \;\; \mathrm{Pr}\left[X_1^n \in \mathcal{S}_n\right] = \epsilon -\mu.
\end{equation}
\underline{Transmitter 1}: Assume it observes the sequence $X_1^n = x_1^n$.
If $x_1^n\notin \mathcal{S}_n$, it looks for  indices \mw{$m_1\geq 1$} satisfying $\left(u_1^{n}(m_1),x^n\right) \in \mathcal{T}_{\mu}^{n}(P_{U_1X_1})$,  randomly picks one of these indices, and sends its corresponding bit-string $\M_1 = \mathrm{string} (m_1)$ both to Transmitter~2 and the Receiver. Otherwise, it sends the single-bit string $\M_1 = [0]$.\\
\underline{Transmitter 2}: Assume it observes the sequence $X_2^n = x_2^n$ and receives \mw{the bit-string}  message $\M_1=\m_1$ from Transmitter 1.
If $\m_1 = [0]$, then it sends \mw{the bit-string} message $\M_2 = [0]$.
Else, if \mw{$m_1=\textnormal{dec}(\m_1)\geq1$, it looks for an index $m_2\geq 1$} satisfying $\left(u_1^{n}(m_1),u_2^{n}(m_2|m_1),x_2^n\right)\in\mathcal{T}_{\mu}^{n}(P_{U_1U_2X_2})$. It randomly picks one of these indices and sends its corresponding bit-string $\M_2 = \mathrm{string} (m_2)$ to the Receiver. Otherwise, it sends $\M_2=[0]$.\\
\underline{Receiver}: Assume it observes the sequence $Y^n = y^n$ and receives messages $\M_1=\m_1$ and $\M_2=\m_2$.
If any of \mw{the bit-strings $\m_1$ or $\m_2$ equals $[0]$,  it declares $\hat{\mathcal{H}} = 1$.
Else, \mw{it sets $m_i=\textnormal{dec}(\m_i)$, for $i=1,2$, and} checks if $\left(u_1^{n}(m_1),u_2^{n}(m_2|m_1),y^n\right)\in\mathcal{T}_{\mu}^{n}(P_{U_1U_2Y})$. It declares $\hat{\mathcal{H}} = 0$ if the condition is verified, and  $\hat{\mathcal{H}} = 1$ otherwise.}
\color{black}
\subsection{Analysis}
Notice first that when $X_1^n \notin \mathcal{S}_n$, our variable-length scheme acts like the fixed-length one in \cite{zhao2018distributed}. We denote by $\hat{\mathcal{H}}^{\textnormal{ZL}}$ the hypothesis guessed by the scheme in \cite{zhao2018distributed}.

The type-I error probability of our scheme satisfies
%{\color{green}Please remove all $\backslash$! and break lines where needed. This is not so nice.}
\begin{IEEEeqnarray}{rCl}
\alpha_n &=& \Pr[\hat{\mathcal{H}}=1|\mathcal{H}=0]\\
&=& \Pr[\hat{\mathcal{H}}=1,X_1^n\hspace{-0.8mm}\in \mathcal{S}_{n} |\mathcal{H}=\!0] \nonumber \\ 
& &+ \Pr[\hat{\mathcal{H}}=1,X_1^n\hspace{-0.8mm}\notin \mathcal{S}_{n}|\mathcal{H}=0]\IEEEeqnarraynumspace\\
&=&\Pr[X_1^n\in \mathcal{S}_{n}|\mathcal{H}=0] \nonumber \\ 
& &+ \Pr[\hat{\mathcal{H}}^{\textnormal{ZL}}=1,X_1^n\notin \mathcal{S}_{n}|\mathcal{H}=0] \IEEEeqnarraynumspace\\
&\leq& \epsilon - \mu + \Pr[\hat{\mathcal{H}}^{\textnormal{ZL}}=1|\mathcal{H}=0].
\end{IEEEeqnarray}
Since by \cite{zhao2018distributed}, $\Pr[\hat{\mathcal{H}}^{\textnormal{ZL}}=1|\mathcal{H}=0] \to 0$ as $n \to  \infty$, we conclude that for the proposed scheme: $\lim_{n\to\infty}\alpha_n \leq \epsilon$.

The type-II error probability satisfies:
\begin{IEEEeqnarray}{rCl}
\beta_n &=& \Pr[\hat{\mathcal{H}}=0|\mathcal{H}=1]\\
&=& \Pr[\hat{\mathcal{H}}=0,X_1^n\in \mathcal{S}_{n}|\mathcal{H}=1] \nonumber\\
& &+ \Pr[\hat{\mathcal{H}}=0, X_1^n\notin \mathcal{S}_{n} |\mathcal{H}=1]\\
&=& \Pr[\hat{\mathcal{H}}^{\textnormal{ZL}}=0, X_1^n\notin \mathcal{S}_{n} |\mathcal{H}=1]\\
&\leq& \Pr[\hat{\mathcal{H}}^{\textnormal{ZL}}=0|\mathcal{H}=1]\\
&\leq& 2^{-n\left(I(U_1U_2;Y)+\delta(\mu)\right)}, \label{BetaZL}
\end{IEEEeqnarray}
where \eqref{BetaZL} uses the achievability result in \cite{zhao2018distributed} and $\delta(\mu) \to 0$ as $\mu \to 0$. Therefore, our scheme achieves the type-II error exponent
\begin{equation}
\theta \geq  I(U_1U_2;Y) + \delta(\mu).
\end{equation}\\
Define $L_1 \triangleq \mathrm{len}(\M_1)$ and $L_2 \triangleq \mathrm{len}(\M_2)$. Notice that \mw{for sufficiently large blocklengths $n$ and $\mu>0$:}
\begin{IEEEeqnarray}{rCl}
\mathbb{E}[L_1] &=& \mathbb{E}[L_1|X_1^n\in \mathcal{S}_n]\Pr[X_1^n\in \mathcal{S}_n]\nonumber\\
& &+\> \mathbb{E}[L_1|X_1^n\notin \mathcal{S}_n]\Pr[X_1^n\notin \mathcal{S}_n]\\
&\leq& (\epsilon - \mu) + n\left(I(U_1;X_1) + \mu\right) \cdot (1 - \epsilon + \mu)\IEEEeqnarraynumspace\\
%& = \epsilon-\mu + n(1 - \epsilon + \mu )I(U_1;X_1) + n\mu(1 - \epsilon + \mu)\\
%& = [\epsilon -\mu -n(\mu\epsilon -\mu^2)] + n[(1 - \epsilon + \mu)I(U_1;X_1) + \mu]\\
&\leq& n(1 - \epsilon + \mu)\left(I(U_1;X_1) + 2\mu\right) \label{EL1_nlarge}\\
&=& nR_1.
\end{IEEEeqnarray}
Similarly, \mw{for sufficiently large blocklengths $n$ and $\mu>0$:} 
\begin{IEEEeqnarray}{rCl}
\mathbb{E}[L_2] &=& \mathbb{E}[L_2|X_1^n\in \mathcal{S}_n ]\Pr[X_1^n\in \mathcal{S}_n ] \\ 
& &+\> \mathbb{E}[L_2|X_1^n\notin \mathcal{S}_n]\Pr[X_1^n\notin \mathcal{S}_n]\\
&\leq& (\epsilon-\mu) + n\left(I(U_2;X_2|U_1) + \mu\right) \cdot (1 - \epsilon + \mu) \IEEEeqnarraynumspace\\
%& = (\epsilon-\mu) + n[(1-\epsilon + \mu)I(U_2;X_2|U_1) + \mu(1-\epsilon + \mu)]\\
%\begin{split}
%& = (\epsilon-\mu) -n\mu(\epsilon - \mu) \\&+ n[(1-\epsilon + \mu)I(U_2;X_2|U_1) + \mu]
%\end{split}\\
&=& \mw{n\left(I(U_2;X_2|U_1) + 2\mu\right) \cdot (1 - \epsilon + \mu) } \label{EL2_nlarge}\\
&\leq& n R_2.
\end{IEEEeqnarray}
\mw{Letting $n\to \infty$ and $\mu \to 0$ concludes our achievability proof.} 
\color{black}
\hfill$\blacksquare$\vspace{0.2cm}

\section{Converse Proof to Theorem~\ref{thm1}}\label{converse}
Notice first that it suffices to show 
\begin{equation}\label{eq:whattoprove}
\theta_{\epsilon}^{*}\left(R_1,R_2\right) \leq \max\limits_{\substack{p\left(u_1|x_1\right)p\left(u_2|u_1,x_2\right):\\R_1 \geq (1-\epsilon)I\left(U_1;X_1\right)\\R_2 \geq (1-\epsilon)I\left(U_2;X_2|U_1\right)\\U_1\leftrightarrow X_1\leftrightarrow (X_2,Y)\\U_2\leftrightarrow (X_2,U_1)\leftrightarrow Y}} I\left(U_1U_2;Y\right),
\end{equation}
i.e., the  Markov chain $U_2 \leftrightarrow (U_1,X_2) \leftrightarrow (X_1,Y)$ in Theorem~\ref{thm1} can be replaced by the weaker Markov chain $U_2 \leftrightarrow (U_1,X_2) \leftrightarrow Y$, because the right-hand side of \eqref{eq:whattoprove} does not depend on the joint pmf of $U_2$ and $X_1$.
\color{black}More formally, we \mw{can} prove the equivalence
\begin{IEEEeqnarray}{rCl}
	& &\bigcup_{\substack{U_1,U_2:\\U_1\leftrightarrow X_1 \leftrightarrow (X_2,Y)\\U_2\leftrightarrow(U_1,X_2)\leftrightarrow(Y,X_1)}}\hspace{-6mm} (I(U_1U_2;Y),I(U_1;X_1),I(U_2;X_2|U_1)) \nonumber\\
	&&\hspace{5mm}= \bigcup_{\substack{U_1,U_2:\\U_1\leftrightarrow X_1 \leftrightarrow (X_2,Y)\\U_2\leftrightarrow(U_1,X_2)\leftrightarrow Y}}\hspace{-6mm} (I(U_1U_2;Y),I(U_1;X_1),I(U_2;X_2|U_1)). \IEEEeqnarraynumspace \label{conversesimplification}
\end{IEEEeqnarray}
Since the two objective functions coincide and the constraints on the \mw{left-hand side (LHS)} are more stringent, it suffices to show that the  \mw{right-hand side (RHS)} is included in the LHS. To this end, fix $U_1,U_2$ satisfying the constraints on the LHS, i.e., the Markov chains $U_1\leftrightarrow X_1 \leftrightarrow (X_2,Y)$ and $U_2\leftrightarrow(U_1,X_2)\leftrightarrow Y$.
Then, construct $\tilde{U}_1,\tilde{U}_2$ so that
\begin{IEEEeqnarray}{rCl}
	P_{\tilde{U}_1|X_1X_2Y}(u_1|x_1,x_2,y) &=& P_{U_1|X_1}(u_1|x_1) \\
	P_{\tilde{U}_2|\tilde{U}_1X_1X_2Y}(u_2|u_1,x_1,x_2,y) &=& P_{U_2|U_1X_2}(u_2|u_1,x_2), \IEEEeqnarraynumspace
\end{IEEEeqnarray}
and thus satisfying the Markov chains on the RHS:\\ $\tilde{U}_1\leftrightarrow X_1 \leftrightarrow (X_2,Y)$ and $\tilde{U}_2\leftrightarrow(\tilde{U}_1,X_2)\leftrightarrow(Y,X_1)$.\\
The proof is concluded by noting that
\begin{align}
	I(\tilde{U}_1;X_1) &= I(U_1;X_1),\label{construction1}\\
	I(\tilde{U}_2;X_2|\tilde{U}_1) &= I(U_2;X_2|U_1),\label{construction2}\\
	I(\tilde{U}_1\tilde{U}_2;Y) &= I(U_1U_2;Y)\label{endofequiv}.
\end{align}
Equalities \eqref{construction1} and \eqref{construction2} hold trivially by construction. Equality~\eqref{endofequiv} holds because $P_{\tilde{U}_1X_2} = P_{U_1X_2}$ and $P_{\tilde{U}_2Y|\tilde{U}_1X_2}=P_{U_2|U_1X_2}\cdot P_{Y|X_2} = P_{U_2Y|U_1X_2}$. 

%We proceed to show that \eqref{eq:whattoprove} holds. Then by \eqref{conversesimplification}, this establishes the desired converse.
\color{black}
%Our proof relies on the following lemma, which is stated without proof.
% \begin{lemma}\label{lem:weak_converse}
%	Let $Q$ and $P$ be arbitrary pmfs over a discrete and finite set $\mathcal{W}$ and $\mathcal{C}$ be a subset of $\mathcal{W}$. Then, 
%	\begin{equation}
%	-\log Q(\mathcal{C})\leq \frac{1}{P(\mathcal{C})} ( D(P \|Q)+ 1).
%	\end{equation}
%\end{lemma}
%\color{black} Do we need to add a proof here since we are in the long version?\\

\color{black} {\color{black}We proceed to show that \eqref{eq:whattoprove} holds.}	Fix $\theta < \theta_{\epsilon}^{*}(R_1,R_2)$, a sequence of encoding and decision functions satisfying the type-I and type-II error constraints, a blocklength $n$, and a small number $\eta \geq 0$.
	Define:
	\begin{IEEEeqnarray}{rCl}\label{Bn}
	\mathcal{B}_n(\eta) &\triangleq& \{(x_1^n,x_2^n)      : \nonumber\\ & &\;\;\mathrm{Pr}[\hat{\mathcal{H}}=0 | X_1^n = x_1^n, X_2^n = x_2^n, \mathcal{H}=0] \geq \eta\}, \IEEEeqnarraynumspace\\
    \mu_n &\triangleq& n^{-{1\over3}},\\
    \mathcal{D}_n(\eta) &\triangleq& T_{\mu_n}^{n}(P_{X_1X_2}) \cap \mathcal{B}_n(\eta).\label{Dn}
	\end{IEEEeqnarray}
	\color{black}By constraint (\ref{type1constraint}) on the type-I error probability:
	\begin{IEEEeqnarray}{rCl}
	1 - \epsilon  &\leq&  \sum_{x_1^{n},x_2^{n}}\Pr[\hat{\mathcal{H}}=0|X_1^n=x_1^n,X_2^n = x_2^n,\mathcal{H}=0]\nonumber\\  
	& & \;\;\;\;\;\;\;\;\;\;\;\;\;\;\;\;\;\;\;\;\;\;\;\;\;\;\;\;\;\;\;\;\;\;\;\;\;\;\;\;\;\;\;\cdot P_{X_1^nX_2^n}(x_1^n,x_2^n)
	\\
	&\leq&\sum_{(x_1^n,x_2^n) \in \mathcal{B}_n(\eta)} P_{X_1^nX_2^n}(x_1^n,x_2^n)\nonumber\\
	&&+\hspace{-3mm}\sum_{(x_1^n,x_2^n) \notin \mathcal{B}_n(\eta)}\hspace{-3mm}\Pr[\hat{\mathcal{H}}=0|X_1^n=x_1^n,X_2^n = x_2^n,\mathcal{H}=0] \nonumber\\
	& &\;\;\;\;\;\;\;\;\;\;\;\;\;\;\;\;\;\;\;\;\;\;\;\;\;\;\;\;\;\;\;\;\;\;\;\;\;\;\;\;\;\;\;\cdot P_{X_1^nX_2^n}(x_1^n,x_2^n)
	\\
	&\leq& P_{X_1^nX_2^n}(\mathcal{B}_n(\eta)) + \eta(1- P_{X_1^nX_2^n}(\mathcal{B}_n(\eta))).
	\end{IEEEeqnarray}
	\color{black}Thus we have:
	\begin{equation}\label{Bnprob}
	P_{X_1^nX_2^n}(B_n(\eta)) \geq {1 - \epsilon - \eta \over 1 - \eta}.
	\end{equation}
	Moreover, by \cite[Lemma~2.12]{Csiszarbook}, \mw{the probability that $(X_1^n,X_2^n)$ lie in the jointly strongly typical set $\mathcal{T}_{\mu_n}^{(n)}(P_{X_1X_2})$ satisfies}
	\begin{equation}\label{Tx1}
	P_{X_1X_2}^{n}\left(\mathcal{T}_{\mu_n}^{(n)}(P_{X_1X_2})\right) \geq 1 - {\vert{\mathcal{X}_1}\vert\ \vert{\mathcal{X}_2}\vert\over{2 \mu_n n}},
	\end{equation}
	\color{black}and since for any events $A$ and $B$,
	\begin{equation}
	\Pr(A\cap B) \geq \Pr(A) + \Pr(B) - 1,
	\end{equation}
	\color{black}then by (\ref{Dn}), \eqref{Bnprob} and (\ref{Tx1}), we obtain
	\begin{equation}\label{Dnprob}
	P_{X_1^nX_2^n}(D_n(\eta)) \geq {1 - \epsilon - \eta \over 1 - \eta} - {\vert \mathcal{X}_1 \vert \vert \mathcal{X}_2 \vert \over {2 \mu_nn}} \triangleq \Delta_n. 
	\end{equation}
	We  define the random variables $(\tilde{\M}_1,\tilde{\M}_2,\tilde{X}_1^n,\tilde{X}_2^n,\tilde{Y}^n)$ as the restriction of the random variables $({\M_1},{\M_2},{X_1^n},{X_2^n},{Y^n})$ to $(X_1^n,X_2^n) \in D_n(\eta)$. The probability distribution of the former tuple is given by:
	\begin{IEEEeqnarray}{rCl}
	&P&_{\tilde{\M}_1\tilde{\M}_2\tilde{X}_1^n\tilde{X}_2^n\tilde{Y}^n}(\m_1,\m_2,x_1^n,x_2^n,y^n) \triangleq \nonumber
	\\ & &\;\;\;\;\;P_{X_1^nX_2^nY^n}(x_1^n,x_2^n,y^n)\cdot{\mathbbm{1} \{x_1^n,x_2^n\in D_n(\eta)\} \over P_{X_1^nX_2^n}(D_n(\eta))} \nonumber
	\\& &\;\;\;\;\;\;\;\;\;\;\cdot{\mathbbm{1}\{\phi_1(x_1^n)=\m_1\}}
	\cdot{\mathbbm{1}\{\phi_2(x_2^n,\phi_1(x_1^n))=\m_2\}}, \IEEEeqnarraynumspace \label{pmftilde}
	\end{IEEEeqnarray}
leading to the following inequalities:
	\begin{equation}\label{tildex1x2relation}
	P_{\tilde{X}_1^n\tilde{X}_2^n}(x_1^n,x_2^n) \leq P_{X_1X_2}^{n}(x_1^n,x_2^n) \Delta_{n}^{-1},
	\end{equation}
	\begin{equation}\label{tildem1m2relation}
	P_{\tilde{\M}_1\tilde{\M}_2}(\m_1,\m_2) \leq P_{\M_1\M_2}(\m_1,\m_2) \Delta_{n}^{-1},
	\end{equation}
	\begin{equation}\label{tildeyrelation}
	P_{\tilde{Y}^n}(y^n) \leq P_{Y}^{n}(y^n) \Delta_n^{-1},
	\end{equation}
	\begin{equation}\label{tildedivergencerelation}
	D(P_{\tilde{X}_1^n\tilde{X}_2^n}||P_{X_1X_2}^{n}) \leq \log{\Delta_n^{-1}}.
	\end{equation}
	
\color{black}\subsection{Single-Letter Characterization of Rate Constraints}
\color{black}Define the following random variables:
\begin{equation}
\tilde{L}_i \triangleq \mathrm{Len}(\tilde{\M}_i),\;\;\;\;\;\; i=1,2.
\end{equation}	
By the rate constraints \eqref{eq:Rate1} and \eqref{eq:Rate2}, we have for $ i =1,2$:
\begin{align}
nR_i &\geq \mathbb{E}[L_i] \label{converseremarkchangebegin}\\
&\geq \mathbb{E}[L_i|(X_1^n,X_2^n) \in D_n(\eta)]P_{X_1^nX_2^n}(D_n(\eta))\\
&\color{black}= \mathbb{E}[\tilde{L}_i]P_{X_1^nX_2^n}(D_n(\eta))\\
&\geq \mathbb{E}[\tilde{L}_i] \Delta_n, \label{ELi}
\end{align}
where the last inequality follows by (\ref{Dnprob}).	
Moreover, by definition, $\tilde{L}_i$ is a function of $\tilde{\M}_i$, for $i=1,2$, so we can upper bound the entropy of $\tilde{\M}_i$ as follows:
\begin{align}
H(\tilde{\M}_i) &= H(\tilde{\M}_i,\tilde{L}_i)\\
&=  H(\tilde{\M}_i|\tilde{L}_i) + H(\tilde{L}_i)\\
&\color{black}= \sum_{l_i} \Pr[\tilde{L}_i = l_i]H(\tilde{\M}_i|\tilde{L}_i=l_i) + H(\tilde{L_i})\\
&\leq \sum_{l_i} \Pr[\tilde{L}_i = l_i]l_i + H(\tilde{L}_i) \label{HMi_ineq1}\\
& = \mathbb{E}[\tilde{L}_i] + H(\tilde{L}_i)\\
&\color{black}\leq {nR_i\over \Delta_n} + H(\tilde{L}_i) \label{HMi_ineq2}\\
&\leq {nR_i\over \Delta_n} + {nR_i\over \Delta_n}{h_b\left({\Delta_n \over nR_i}\right)} \label{HMi_ineq3}\\
&= {nR_i\over \Delta_n} \left(1 + {h_b\left({\Delta_n \over nR_i}\right)}\right), \label{Miub}
\end{align}
where \eqref{HMi_ineq2} holds by (\ref{ELi}), and \eqref{HMi_ineq3} holds since the maximum possible entropy of $\tilde{L}_i$ is obtained by a geometric distribution of mean $\mathbb{E}[\tilde{L}_i]$, which is further bounded by $nR_i \over \Delta_n$ \cite[Theorem 12.1.1]{cover}.\\

On the other hand, we lower bound the entropy of $\tilde{\M}_1$ as:
\begin{IEEEeqnarray}{rCl}
H(\tilde{\M}_1)\!&\geq&\! I(\tilde{\M}_1;\tilde{X}_1^n\tilde{X}_2^n) + D(P_{\tilde{X}_1^n\tilde{X}_2^n}||P_{X_1X_2}^n) + \log\Delta_{n}\IEEEeqnarraynumspace\label{m1entropylbstep1}\\
&=&\! H(\tilde{X}_1^n\tilde{X}_2^n) + D(P_{\tilde{X}_1^n\tilde{X}_2^n}||P_{X_1X_2}^n) \nonumber\\& &-  H(\tilde{X}_1^n\tilde{X}_2^n|\tilde{\M}_1) + \log\Delta_{n}\\
&\geq&\color{black}\! n [H(\tilde{X}_{1,T}\tilde{X}_{2,T}) + D(P_{\tilde{X}_{1,T}\tilde{X}_{2,T}}||P_{X_1X_2})] \nonumber\\
& &\color{black}- \sum_{t=1}^{n} H(\tilde{X}_{1,t}\tilde{X}_{2,t}|\tilde{\M}_1\tilde{X}_1^{t-1}\tilde{X}_2^{t-1}) + \log\Delta_{n}\label{m1entropylbstep2}\\
&=&\! n [H(\tilde{X}_{1,T}\tilde{X}_{2,T}) + D(P_{\tilde{X}_{1,T}\tilde{X}_{2,T}}||P_{X_1X_2})] \nonumber\\
& &- \sum_{t=1}^{n} H(\tilde{X}_{1,t}\tilde{X}_{2,t}|\tilde{U}_{1,t})+ \log\Delta_{n}\label{m1entropylbstep4}\\
&=&\! n [H(\tilde{X}_{1,T}\tilde{X}_{2,T}) + D(P_{\tilde{X}_{1,T}\tilde{X}_{2,T}}||P_{X_1X_2})] \nonumber\\
& &- n H(\tilde{X}_{1,T}\tilde{X}_{2,T}|\tilde{U}_{1,T},T)+ \log\Delta_{n} \label{Tuniformdef}\\
&=&\! n [H(\tilde{X}_{1}\tilde{X}_{2}) + D(P_{\tilde{X}_{1}\tilde{X}_{2}}||P_{X_1X_2})] \nonumber\\
& &- n H(\tilde{X}_{1}\tilde{X}_{2}|{U}_{1})+ \log\Delta_{n}\label{m1entropylbstep5}\\
&=&\color{black}\! n\left[I(U_1;\tilde{X}_{1}\tilde{X}_{2})\!+\! D(P_{\tilde{X}_{1}\tilde{X}_{2}}||P_{X_1X_2})\right]\!+\!\log\Delta_{n} \IEEEeqnarraynumspace\\
&\geq&\! n \left[I(U_1;\tilde{X}_{1}) + {1 \over n}\log\Delta_{n}\right].\label{M1lb}
\end{IEEEeqnarray}
Here, (\ref{m1entropylbstep1}) holds by (\ref{tildedivergencerelation}); (\ref{m1entropylbstep2}) holds by the super-additivity property in \cite[Proposition 1]{tyagi2019strong} and by the chain rule; \eqref{m1entropylbstep4} holds by defining $\tilde{U}_{1t} \triangleq (\tilde{\M}_1,\tilde{X}_{1}^{t-1},\tilde{X}_2^{t-1})$; \eqref{Tuniformdef} holds by defining $T$ uniform over $\{1,\dots,n\}$ independent of all other random variables; and (\ref{m1entropylbstep5}) holds by defining $U_1 \triangleq (\tilde{U}_{1T},T)$, $\tilde{X}_1\triangleq \tilde{X}_{1,T}$, and $\tilde{X}_2\triangleq \tilde{X}_{2,T}$.

Similarly,
\begin{IEEEeqnarray}{rCl}
H(\tilde{\M}_2) &\geq& I(\tilde{\M}_2;\tilde{X}_1^n\tilde{X}_2^n|\tilde{\M}_1)\label{m2entropylbstep1}\\
&=&\color{black}\sum_{t=1}^{n}I(\tilde{\M}_2;\tilde{X}_{1,t}\tilde{X}_{2,t}|\tilde{\M}_1\tilde{X}_1^{t-1}\tilde{X}_2^{t-1})\label{m2entropylbstep2}\\
&=&\sum_{t=1}^{n}I(\tilde{U}_{2,t};\tilde{X}_{1,t}\tilde{X}_{2,t}|\tilde{U}_{1,t})\label{m2entropylbstep3}\\
&=&nI(\tilde{U}_{2,T};\tilde{X}_{1,T}\tilde{X}_{2,T}|\tilde{U}_{1,T}T)\\
&=&\color{black}nI(\tilde{U}_{2,T}T;\tilde{X}_{1,T}\tilde{X}_{2,T}|\tilde{U}_{1,T}T)\\
&=&\color{black}nI({U}_{2};\tilde{X}_{1}\tilde{X}_{2}|{U}_{1})\label{m2entropylbstep4}\\
&\geq& nI({U}_{2};\tilde{X}_{2}|{U}_{1})\label{M2lb}.
\end{IEEEeqnarray}
Here, (\ref{m2entropylbstep1}) holds since $\tilde{\M}_2$ is function of $\tilde{X}_2^n$ and $\tilde{\M}_1$; \eqref{m2entropylbstep2} holds by the chain rule; (\ref{m2entropylbstep3}) holds by the definition of $\tilde{U}_{1t}$, and by defining $\tilde{U}_{2t} \triangleq \tilde{\M}_2$; and (\ref{M2lb}) holds by defining $U_2 \triangleq (\tilde{U}_{2T},T)$.

Combining (\ref{Miub}) with (\ref{M1lb}) and (\ref{M2lb}) yields:
\begin{IEEEeqnarray}{rCl}
R_1 &\geq & {I(U_1;\tilde{X}_1) + {1 \over n} \log \Delta_{n} \over \left(1 + {h_b\left({\Delta_n \over nR_1}\right)}\right)}\cdot \Delta_n \label{R1lb} \\
R_2 &\geq &{I(U_2;\tilde{X}_2|U_1) \over \left(1 + {h_b\left({\Delta_n \over nR_2}\right)}\right)}\cdot \Delta_n. \label{R2lb}
\end{IEEEeqnarray}

\color{black}\subsection{Upper Bounding the Type-II Error Exponent}
\color{black}Define for each $(\m_1,\m_2)$ the set
\begin{equation}
\mathcal{A}_n(\m_1,\m_2) \triangleq \{y^n \colon (\m_1,\m_2,y^n) \in \mathcal{A}_n\},
\end{equation}
and its Hamming neighborhood:
\begin{IEEEeqnarray}{rCl}
\hat{\mathcal{A}}_n^{\ell_n}(\m_1,\m_2) \triangleq \{\tilde{y}^n : \exists \, y^n \in &&\,\mathcal{A}_n(\m_1,\m_2) \nonumber\\ &&\; \textnormal{s.t.} \; d_H(y^n,\tilde{y}^n)\leq\ell_n\}
\end{IEEEeqnarray}
for some real number $\ell_n$ satisfying $\lim_{n \rightarrow \infty} {\ell_n/n} =0 $ and \color{black} $\lim_{n \rightarrow \infty} {\ell_n/\sqrt{n}} =\infty $.\\
%\begin{equation}
%\ell_n \geq \sqrt{{n \over 2} \log{1 \over 1-\epsilon}}.
%\end{equation}
\color{black}Since by definitions (\ref{Bn}) and (\ref{Dn}), for all $(x_1^n,x_2^n) \in \mathcal{D}_n$:
\begin{equation}\label{blowupcond}
P_{\tilde{Y}^n|\tilde{X}_1^n\tilde{X}_2^n}(\mathcal{A}_n(\m_1,\m_2)|x_1^n,x_2^n) \geq \eta ,
\end{equation}
then by the blowing-up lemma \cite{MartonBU}:
\begin{equation}\label{blowup1}
P_{\tilde{Y}^n|\tilde{X}_1^n\tilde{X}_2^n}(\hat{\mathcal{A}}_n^{\ell_n}(\m_1,\m_2)|x_1^n,x_2^n) \geq 1 - \zeta_n
\end{equation}
for a real number $\zeta_n > 0$ such that $\lim_{n \to \infty} \zeta_n = 0$. Moreover, by taking the expectation over \eqref{blowup1}:

\color{black}\begin{IEEEeqnarray}{rCl}
P_{\tilde{\M}_1\tilde{\M}_2\tilde{Y}^n}(\hat{\mathcal{A}}_n^{\ell_n}) &=&
\hspace{-5mm}\sum_{\substack{(x_1^n,x_2^n)\in \mathcal{D}_n\\(\m_1,\m_2)\in\mathcal{M}_1\times\mathcal{M}_2}}\hspace{-6mm} P_{\tilde{Y}^n|\tilde{X}_1^n\tilde{X}_2^n}(\hat{\mathcal{A}}_n^{\ell_n}(\m_1,\m_2)|x_1^n,x_2^n)\nonumber\\ 
&&\hspace{1.73cm} \cdot P_{\tilde{X}_1^n\tilde{X}_2^n\tilde{\M}_1\tilde{\M}_2}(x_1^n,x_2^n,\m_1,\m_2) \nonumber \\
& \geq &1 - \zeta_n. \IEEEeqnarraynumspace
\end{IEEEeqnarray}
\color{black}
In addition, using (\ref{tildem1m2relation}) and (\ref{tildeyrelation}), we have the following:
\begin{IEEEeqnarray}{rCl}\label{betanewub1}
P_{\tilde{\M}_1\tilde{\M}_2}&P_{\tilde{Y}^n}&(\hat{\mathcal{A}}_n^{\ell_n}) \nonumber\\
&\leq& P_{\M_1 \M_2} P^{n}_{Y}(\hat{\mathcal{A}}_n^{\ell_n}) \cdot \Delta_n^{-2}\\
&\leq& P_{\M_1 \M_2} P^{n}_{Y}\left({\mathcal{A}_n}\right) \cdot e^{nh_b(\ell_n/n)} \cdot p^{\ell_n} \cdot {\vert \mathcal{Y} \vert}^{\ell_n} \cdot \Delta_n^{-2}\IEEEeqnarraynumspace\label{Eq:ByCsiszarKornerLemma}\\
%&=& \beta_n\cdot e^{nh_b(\ell_n/n)} \cdot p^{\ell_n} \cdot {\vert \mathcal{Y} \vert}^{\ell_n} \cdot \Delta_n^{-2}\\
&=& \beta_n\cdot F_n^{\ell_n}\cdot \Delta_n^{-2},\label{Eq:betaandQrelation}
\end{IEEEeqnarray}
where $p \triangleq \min\limits_{\substack{y,y':P_Y(y') > 0}}{P_Y(y) \over P_Y(y')}$ and $ F_n^{\ell_n} \triangleq e^{nh_b(\ell_n/n)} \cdot p^{\ell_n} \cdot {\vert \mathcal{Y} \vert}^{\ell_n}$. Here, (\ref{Eq:ByCsiszarKornerLemma}) holds by \cite[Proof of Lemma 5.1]{Csiszarbook}.
%This can be further bounded as follows:
%
%\begin{align}
%P_{M_1 M_2} P^{n}_{Y}\left(\widehat{{\mathcal{A}}_n^{\ell_n}}\right) &= \Pr\left[(M_1,M_2,Y^n)\in\widehat{{\mathcal{A}}_n^{\ell_n}} | H= H_1\right]\\
%&\leq \sum_{(m_1,m_2,y^n)\in \mathcal{A}_n}\;\sum_{\bar{y}^n \in \mathrm{Ball}_{\ell_n}(y^n)}P_Y^n(\bar{y}^n)\cdot  P_{M_1 M_2}({m}_1 {m}_2)\\
%&\begin{multlined}= \sum_{(m_1,m_2,y^n)\in \mathcal{A}_n}\;\sum_{\bar{y}^n \in  \mathrm{Ball}_{\ell_n}(y^n)}{P_Y^n(\bar{y}^n) \over P_Y^n(y^n)} \cdot { P_Y^n(y^n)P_{M_1 M_2}(m_1m_2)}\;\;\;\;\end{multlined}\\
%&\leq \sum_{(m_1,m_2,y^n)\in \mathcal{A}_n}\;\sum_{\bar{y}^n \in \mathrm{Ball}_{\ell_n}(y^n)} p^{\ell_n} \cdot { P_Y^n(y^n)P_{M_1 M_2}(m_1m_2)}\\
%&\begin{multlined}\leq \sum_{(m_1,m_2,y^n)\in \mathcal{A}_n} {n \choose \ell_n} \cdot p^{\ell_n} \cdot {\vert \mathcal{Y} \vert}^{\ell_n} \cdot { P_Y^n(y^n)P_{M_1 M_2}(m_1m_2)}\end{multlined}\\
%&\begin{multlined}= {n \choose \ell_n} \cdot p^{\ell_n} \cdot {\vert \mathcal{Y} \vert}^{\ell_n} \cdot \Pr \left[(M_1, M_2, Y^n)\in {\mathcal{A}}_n | H=H_1 \right] \end{multlined}\\
%&\leq {\left( {ne \over \ell_n} p {\vert \mathcal{Y} \vert}\right)}^{\ell_n}\cdot \beta_n\\
%&= F_n^{\ell_n} \cdot \beta_n \label{Fnbetan}
%\end{align}
%Where $p \triangleq \max\limits_{\substack{y,y':\\P_Y(y') > 0}}{P_Y(y) \over P_Y(y')}$.\newline

By (\ref{Eq:betaandQrelation}) and standard inequalities (see \cite[Lemma~1]{MicheleVL}), we can upper bound the type-II error exponent as follows:
\begin{IEEEeqnarray}{rCl}
\lefteqn{-{\log{\beta_n}}} \nonumber\\  
&\leq& -{\log{P_{\tilde{\M}_1\tilde{\M}_2}P_{\tilde{Y}^n}}}(\hat{\mathcal{A}}_n^{\ell_n}) + {\ell_n}\log F_n - 2\log{\Delta_n}\\
&\leq& {1 \over 1-\zeta_n}\left( D\left(P_{\tilde{\M}_1\tilde{\M}_2\tilde{Y}^n} || P_{\tilde{\M}_1\tilde{\M}_2}P_{\tilde{Y}^n}\right) + 1\right)\nonumber\\ 
&& +\> {\ell_n}\log F_n - {2}\log \Delta_n\\
&=& {1 \over 1 - \zeta_n}\left(I({\tilde{\M}_1\tilde{\M}_2;\tilde{Y}^n})+ 1\right) + {\ell_n}\log F_n - {2}\log \Delta_n. \IEEEeqnarraynumspace \label{thetaub2}
\end{IEEEeqnarray}
We further upper-bound the term $I(\tilde{\M}_1\tilde{\M}_2;\tilde{Y}^n)$ as follows:
\color{black}\begin{align}
I(\tilde{\M_1}\tilde{\M_2};\tilde{Y}^n) &= \sum_{t=1}^{n}I(\tilde{\M_1}\tilde{\M_2};\tilde{Y}_t|\tilde{Y}^{t-1})\\
&\leq \sum_{t=1}^{n}I(\tilde{\M_1}\tilde{\M_2}\tilde{X}_1^{t-1}\tilde{X}_2^{t-1}\tilde{Y}^{t-1};\tilde{Y}_t)\\
&= \sum_{t=1}^{n}I(\tilde{\M}_1\tilde{\M}_2\tilde{X}_1^{t-1}\tilde{X}_2^{t-1};\tilde{Y}_t)\label{Markovchain2}\\
&= \sum_{t=1}^{n}I(\tilde{U}_{1,t}\tilde{U}_{2,t};\tilde{Y}_t)\label{}\label{U1U2tdef}\\
&= n I(\tilde{U}_{1,T}\tilde{U}_{2,T};\tilde{Y}_T|T)\\
&\leq  n I({U}_{1}{U}_{2};\tilde{Y}),\label{thetaub}
\end{align}\color{black}
where (\ref{Markovchain2}) holds by the Markov chain $\tilde{Y}^{t-1} \leftrightarrow (\tilde{\M}_1\tilde{\M}_2,\tilde{X}_1^{t-1}\tilde{X}_2^{t-1}) \leftrightarrow \tilde{Y}_t$; (\ref{thetaub}) follows by the definitions of \mw{$\tilde{U}_{1}$ and $\tilde{U}_{2}$} and defining $\tilde{Y} = \tilde{Y}_T$.

{\color{black}\subsection{Establishing the Desired Markov Chains}}
We observe the Markov chain $\tilde{U}_{2,t} \leftrightarrow (\tilde{U}_{1,t},\tilde{X}_{2,t}) \leftrightarrow \tilde{Y}_t$  for any $t$, \mw{and thus $U_2 \leftrightarrow (U_1,\tilde{X}_2)\leftrightarrow \tilde{Y}$.}  The second desired Markov chain \mw{${U}_{1} \leftrightarrow \tilde{X}_{1} \leftrightarrow (\tilde{X}_{2},\tilde{Y})$} only holds in the limit as $n \to \infty$.
%\begin{equation}\label{Markovchainforninfinity}
%\lim_{n \to \infty} {1 \over n} [I(\tilde{\M}_1;\tilde{X}_2^n\tilde{Y}^n|\tilde{X}_1^n) + D(P_{\tilde{X}_1^n\tilde{X}_2^n\tilde{Y}^n}||P_{X_1X_2Y}^n)] = 0
%\end{equation}
\mw{To see this}, notice that $\tilde{\M}_1 \leftrightarrow \tilde{X}_1^n \leftrightarrow (\tilde{X}_2^n,\tilde{Y}^n)$ forms a Markov chain and thus:
\begin{IEEEeqnarray}{rCl}
0\! &=&\!I(\tilde{\M}_1;\tilde{X}_2^n\tilde{Y}^n|\tilde{X}_1^n) \\ 
&\geq&\!H(\tilde{X}_2^n\tilde{Y}^n|\tilde{X}_1^n) + D(P_{\tilde{X}_1^n\tilde{X}_2^n\tilde{Y}^n}||P_{X_1X_2Y}^n) \nonumber\\
&& +\log{\Delta_{n}} - H(\tilde{X}_2^n\tilde{Y}^n|\tilde{X}_1^n\tilde{\M}_1)
\label{MC1proofstep1}\\
%&=&\!n[H(\tilde{X}_{2,T}\tilde{Y}_T|\tilde{X}_{1,T}) + D(P_{\tilde{X}_{1,T}\tilde{X}_{2,T}\tilde{Y}_T}||P_{X_1X_2Y})] \nonumber\\
%&&+\log{\Delta_{n}} - H(\tilde{X}_2^n\tilde{Y}^n|\tilde{X}_1^n\tilde{\M}_1)\label{MC1proofstep2}\\
&=&\!n[H(\tilde{X}_{2,T}\tilde{Y}_T|\tilde{X}_{1,T}) + D(P_{\tilde{X}_{1,T}\tilde{X}_{2,T}\tilde{Y}_T}||P_{X_1X_2Y})]\nonumber\\
&&+\log{\Delta_{n}}-\sum_{t=1}^{n}H(\tilde{X}_{2,t}\tilde{Y}_t|\tilde{X}_1^n\tilde{X}_{2}^{t-1}\tilde{Y}^{t-1}\tilde{\M}_1)\label{MC1proofstep2}\\
&\geq&\color{black}\! n[H(\tilde{X}_{2,T}\tilde{Y}_T|\tilde{X}_{1,T}) + D(P_{\tilde{X}_{1,T}\tilde{X}_{2,T}\tilde{Y}_T}||P_{X_1X_2Y})]\nonumber\\
&&\color{black}+\log{\Delta_{n}}-\sum_{t=1}^{n}H(\tilde{X}_{2,t}\tilde{Y}_t|\tilde{X}_1^{t}\tilde{X}_{2}^{t-1}\tilde{\M}_1)\label{MC1proofstep3}\\
&=&\!n[H(\tilde{X}_{2,T}\tilde{Y}_T|\tilde{X}_{1,T}) + D(P_{\tilde{X}_{1,T}\tilde{X}_{2,T}\tilde{Y}_T}||P_{X_1X_2Y})]\nonumber\\ 
&&+\log{\Delta_{n}} -  \sum_{t=1}^{n}H(\tilde{X}_{2,t}\tilde{Y}_t|\tilde{X}_{1,t}\tilde{U}_{1,t})\label{MC1proofstep4}\\
&=&\color{black}\! n[H(\tilde{X}_{2,T}\tilde{Y}_T|\tilde{X}_{1,T}) + D(P_{\tilde{X}_{1,T}\tilde{X}_{2,T}\tilde{Y}_T}||P_{X_1X_2Y})] \nonumber\\
&&\color{black} +\log{\Delta_{n}} -  n H(\tilde{X}_{2,T}\tilde{Y}_T|\tilde{X}_{1,T}\tilde{U}_{1,T}T)\\
&=&\!n[I(\tilde{X}_{2,T}\tilde{Y}_T;\tilde{U}_{1,T}T|\tilde{X}_{1,T}) \nonumber\\
&& + D(P_{\tilde{X}_{1,T}\tilde{X}_{2,T}\tilde{Y}_T}||P_{X_1X_2Y})] +\log{\Delta_{n}}\\
&=&\color{black}\!n[I(\tilde{X}_{2}\tilde{Y};{U}_{1}|\tilde{X}_{1})\!+\! D(P_{\tilde{X}_{1}\tilde{X}_{2}\tilde{Y}}||P_{X_1X_2Y})]\!+\!\log{\Delta_{n}} \IEEEeqnarraynumspace \label{MC1proofstep5}\\
&\geq&\!nI(\tilde{X}_{2}\tilde{Y};{U}_{1}|\tilde{X}_{1})\!+ \log{\Delta_{n}},\label{thetaterm3}
\end{IEEEeqnarray}
where \eqref{MC1proofstep1} follows by \eqref{tildedivergencerelation} and $P_{\tilde{Y}^n|\tilde{X}_1^n\tilde{X}_2^n}=P_{Y|X_1X_2}^n$; \eqref{MC1proofstep2} holds by the super-additivity property in \cite[Proposition 1]{tyagi2019strong} and the chain rule; \eqref{MC1proofstep3} holds since knowledge reduces entropy; and finally \eqref{MC1proofstep4} and \eqref{MC1proofstep5} hold by the definitions of $\tilde{U}_{1,t},\tilde{X}_2,\tilde{Y}$, and $U_1$. Moreover, since ${1\over n}\log{\Delta_{n}} \to 0$ as $n \to \infty$, then $I(\tilde{X}_{2}\tilde{Y};{U}_{1}|\tilde{X}_{1}) \to 0$ as $n \to \infty$. % and thus the Markov chain $\tilde{U}_{1} \leftrightarrow \tilde{X}_{1} \leftrightarrow (\tilde{X}_{2},\tilde{Y})$ holds if a convergence of $P_{\tilde{X}_1\tilde{X}_2\tilde{Y}\tilde{U}_1\tilde{U}_2}^{(i_k)}$ can be shown.

{\color{black}\subsection{The Limits $n\to \infty$ and $\eta\to 0$}}
To sum up, we have proved so far in \eqref{R1lb}, \eqref{R2lb}, \eqref{thetaub2}, \eqref{thetaub}, and \eqref{thetaterm3} that for all $n\geq 1$ there exists a joint pmf $P_{\tilde{X}_1\tilde{X}_2\tilde{Y}{U}_1{U}_2}^{(n)}$ (abbreviated as $P^{(n)}$) \mw{and functions $g_1(n)$, $g_3(n)$, and $g_4(n)$ tending to 0 as $n \to \infty$ and  $g_2(n,\eta)$ tending to $(1-\epsilon)$ as $n \to \infty$ and $\eta \to 0$, so that 
\begin{subequations}\label{eq:conditions}
\begin{IEEEeqnarray}{rCl}
P_{\tilde{X}_1\tilde{X}_2\tilde{Y}{U}_1{U}_2}^{(n)}& = &P_{\tilde{X}_1\tilde{X}_2\tilde{Y}}^{(n)}\cdot P_{U_1|\tilde{X}_1\tilde{X}_2}^{(n)}\cdot P_{U_2|U_1\tilde{X}_2}^{(n)},\IEEEeqnarraynumspace\\
R_1& \geq & (I_{P^{(n)}}(U_1;\tilde{X}_1) + g_1(n))\cdot g_2(n,\eta),\nonumber \\ \\
R_2& \geq &I_{P^{(n)}}(U_2;\tilde{X}_2|U_1)\cdot g_2(n,\eta), \\
\theta &\leq& I_{P^{(n)}}(U_1U_2;\tilde{Y}) + g_3(n),\\
I_{P^{(n)}}(\tilde{X}_2\tilde{Y};U_1|\tilde{X}_1)& \leq &g_4(n),\label{eq:last_cond}
\end{IEEEeqnarray}
\end{subequations}
where $I_{P^{(n)}}$ indicates that the mutual information should be calculated according to the pmf $P^{(n)}$.}
%Also note that $P_{\tilde{Y} |\tilde{X}_1\tilde{X}_2} = P_{Y|X_1X_2}$, and  since $\tilde{X}_1^n\tilde{X}_2^n$ are restricted to their typical sets i.e. $\vert P_{\tilde{X}_1\tilde{X}_2} - P_{X_1X_2}\vert \leq \mu_n$, it holds that $P^{(n)}

\mw{Applying} Carath\'eodory's theorem \cite[Appendix C]{ElGamal}, \mw{one can restrict the auxiliary random variables $U_1$ and $U_2$ to alphabets of sizes}
\begin{align}
\vert \mathcal{U}_1\vert &\leq \vert \mathcal{X}_1\vert\cdot\vert \mathcal{X}_2\vert + 3,\\
\vert \mathcal{U}_2\vert &\leq \vert \mathcal{U}_1\vert\cdot\vert \mathcal{X}_2\vert + 1.
\end{align}
\mw{The proof is then concluded by invoking the Bolzano-Weierstrass theorem, and by considering a subsequence $P_{\tilde{X}_1\tilde{X}_2\tilde{Y}U_1U_2}^{(n_k)}$ that converges to a limiting pmf  $P_{X_1X_2YU_1U_2}^{*}$. In fact, by  \eqref{eq:conditions} this  limiting pmf  factorizes as $P_{X_1X_2YU_1U_2}^{*}=P_{X_1X_2Y}^*\cdot P_{U_1|X_1}^{*}\cdot P_{U_2|U_1X_2}^{*}$ and satisfies the desired rate-constraints, and moreover $P_{X_1X_2Y}^*=P_{X_1X_2}\cdot P_{Y|X_1X_2}$ because  for any $n \geq 1$,  $P_{\tilde{Y} |\tilde{X}_1\tilde{X}_2}^{(n)} = P_{Y|X_1X_2}$ and  $\vert P_{\tilde{X}_1\tilde{X}_2} - P_{X_1X_2}\vert \leq \mu_n$ (since ($\tilde{X}_1^n,\tilde{X}_2^n)\in \mathcal{T}^{(n)}_{\mu_n}(P_{X_1X_2})$) with $\mu_n \to 0$ as $n\to \infty$. } \hfill $\blacksquare$

\vspace{0.3cm}

\section*{Acknowledgment}
{M. Wigger and M. Hamad acknowledge funding support from the ERC under grant agreement 715111.}
\vspace{0.2cm}

\bibliographystyle{ieeetr}
\bibliography{references}
%\begin{thebibliography}{00}
%\bibitem{b1}@book{el2011network,
%	title={Network information theory},
%	author={El Gamal, Abbas and Kim, Young-Han},
%	year={2011},
%	publisher={Cambridge university press}
%}
%\end{thebibliography}
\end{document}